\documentclass[singlecolumn,showpacs,preprintnumbers,amsmath,amssymb,superscriptaddress]{revtex4}
\usepackage{graphicx}% Include figure files
\usepackage{dcolumn}% Align table columns on decimal point
\usepackage{bm}% bold math
\usepackage{longtable}% bold math
\usepackage{braket}
\newcommand{\bea}{\begin{eqnarray}}
\newcommand{\eea}{\end{eqnarray}}

\usepackage{amsfonts}
\usepackage{amssymb}
\usepackage{amsmath}
\usepackage{amsthm}
\usepackage{graphicx}

\newcommand{\be}{\begin{equation}}
\newcommand{\ee}{\end{equation}}

\def\Be'{\beta_\mu^{'}}

\def\<{\bigl\langle}
\def\>{\bigr\rangle}

\begin{document}

%\preprint{APS/123-QED}

\title{Complex Reaction Kinetics in Chemistry: \\ A unified picture suggested by Mechanics in Physics}

\author{Elena Agliari}
\affiliation{Dipartimento di Matematica, Sapienza Universit\`a di Roma, GNFM-INdAM Sezione di Roma, Italy}
\author{Adriano Barra}
\affiliation{Dipartimento di Matematica e Fisica {\em Ennio De Giorgi}, Universit\`a del Salento, GNFM-INdAM Sezione di Roma, INFN Sezione di Lecce, Italy}
\author{Giulio Landolfi}
\affiliation{Dipartimento di Matematica e Fisica {\em Ennio De Giorgi}, Universit\`a del Salento, INFN Sezione di Lecce, Italy}
\author{Sara Murciano}
\affiliation{Dipartimento di Matematica e Fisica {\em Ennio De Giorgi}, Universit\`a del Salento, Italy, D\'epartment de Physique, \'Ecole Normale Sup\'eriore, France}
\author{Sarah Perrone}
\affiliation{Dipartimento di Fisica, Universit\`a di Torino, Italy}

\date{\today}

\begin{abstract}
Complex biochemical pathways or regulatory enzyme kinetics can be reduced to chains of elementary reactions, which can be described in terms of chemical kinetics. This discipline provides a set of tools for quantifying and understanding the {\em dialogue} between reactants, whose framing into a solid and consistent mathematical description is of pivotal importance in the growing field of biotechnology. Among the elementary reactions so far extensively investigated, we recall the so-called Michaelis-Menten scheme and the Hill positive-cooperative kinetics, which apply to molecular binding and are characterized by the absence and the presence, respectively, of cooperative interactions between binding sites, giving rise to qualitative different phenomenologies.
%In the former case, no interaction occur among the binding sites, while in the latter case binding sites
However, there is evidence of reactions displaying a more complex, and by far less understood, pattern: these follow the positive-cooperative scenario at small substrate concentration, yet negative-cooperative effects emerge and get stronger as the substrate concentration is increased.
\newline
In this paper we analyze the structural analogy between the mathematical backbone of (classical) reaction kinetics in Chemistry and that of (classical) mechanics in Physics: techniques and results from the latter shall be used to infer properties on the former.  In particular, we first show that standard cooperative kinetics can be framed in terms of classical mechanics, and, interestingly, the emerging phenomenology, usually obtained by applying the thermodynamic principles, can be obtained by applying the principle of least action of classical mechanics. Further, since the saturation function (that is a bounded function) plays in Chemistry the same role played by velocity in Physics, we show how a relativistic scaffold naturally accounts also for the kinetics of the above-mentioned complex reactions. Of course, in the classical limit this generalized theory recovers the correct analytical expressions of standard chemical kinetics.
\newline
The reward in the proposed formalism  is two-fold: a unique, consistent picture for cooperative-like reactions in chemical kinetics and a stronger and robust mathematical control, particularly useful to tackle reactions involving small numbers of molecules as those studied in nowadays experiments.
\end{abstract}

%Keywords: statistical mechanics, spin-glasses, complex networks, theoretical immunology.

%\pacs{05.40.Fb, 02.50.Cw, 02.50.Ey}

%\pacs{87.16.Yc, 02.10.Ox, 87.19.xw, 64.60.De, 84.35.+i}

\maketitle

\section{Introduction}
\subsection{The Chemical Kinetics background} \label{sec:intro}

The mathematical models that describe reaction kinetics provide chemists and chemical engineers with tools to better understand, depict and possibly control a broad range of chemical processes (see e.g., \cite{bookCK,bookMazza}). These include applications to pharmacology, environmental pollution monitoring, food industry, etc.
%
%food decomposition, microorganism growth, stratospheric ozone decomposition, and the complex chemistry of biological systems)These models can also be used in the design or modification of chemical reactors to optimize product yield, more efficiently separate products, and eliminate environmentally harmful by-products.
%
In particular, biological systems are often characterized by complex chemical pathways whose modeling is rather challenging and can not be recast in standard schemes \cite{Agliari2,Agliari4,Complexity3,Angeli,Science1,Crampin,Chen,Complexity1,urka,MicroRNA2,Mazza,Perc1,Perc2} (see also \cite{Complex1,Ricci2,Ricci3,Valant} for a different perspective).
%
%
%
%Chemical kinetics traces its origins back in time \cite{} but it entered in the world of Complexity since it started to cover a pivotal role  in Biochemistry, %ranging from regulating  enzyme kinetics up to coding cascades of complex pathways: the first chemical kinetic paradigm that consolidated along the years %in this field was the Micaelis-Menten scenario\cite{}.
In general, one tries to split such sophisticated systems into a set of elementary constituents, in mutual interaction, and for which a clear formalization is available \cite{Agliari5,SciRepKatz,Katz1,Winfree1,Winfree2,Katz2}.

In this context, one of the best consolidated, elementary scheme is given by the Michaelis-Menten law.
This was originally introduced by Leonor Michaelis and Maud Menten to describe enzyme kinetics and can be applied to systems made of two reactants, say $A$ (the binding molecule or, more generally, the binding sites of a molecule) and $B$ (the free ligand, i.e., the substrate), which can bind (and unbind) to form the product $AB$. If we call $S$ the concentration of free ligand, $Y$ the \emph{saturation function} (or \emph{fractional occupancy}), namely the fraction of bound molecules ($Y \in [0,1]$), and, accordingly, $1-Y$ the fraction of the unbound molecules, under proper assumptions, one can write
\be
S (1- Y ) = k Y,
\ee
where $k$ is the proportionality constant between response and occupancy (otherwise stated, it is the ratio between the dissociation and the association constants). In particular, as standard, it is assumed that the $(a)$ the reaction is in a steady state, with the product being formed and consumed at the same rate, $(b)$ the free ligand concentration is in large excess over that of the binding molecules in such a way that it can be considered as constant along the reaction, $(c)$ all the binding molecules are equivalent and independent. Also, the derivation of the Michaelis-Menten law is based on the law of mass action.
\newline
Reshuffling the previous equation we get $Y =  S/ (S + k)$ which allows stating that $k$ is the concentration of free ligand at which $50 \%$ of the binding sites are occupied (that is, when $S=k$, then $Y=1/2$). Thus, denoting with $S_0$ the half-saturation ligand concentration, we get
\be \label{eq:MM}
Y = \frac{S}{{S} + S_0}.
\ee
%
%Equation \ref{eq:MM}
This equation represents a rectangular hyperbola with horizontal asymptote corresponding to full saturation, that is $Y=1$; this is the typical outcome expected for systems where no interaction between binding sites is at work \cite{Thompson}.
%
%
%Leonor Michaelis and Maud Menten pointed out that the velocity of a reaction $v_r$ was proportional to a simple monomial of the substrate concentration $S$, namely
%\be\label{MMoriginal}
%v_r \propto \frac{S}{S_0+S},
%\ee
%where $S_0$ is the substrate concentration corresponding to half of the maximum of the saturation function. The latter, $Y$, is the percentage of binding sites that have been occupied by the ligand, thus $Y \in [0,1]$.
%\newline
%This observation, coupled to the Law of Mass Action that (oversimplifying) states that the percentage of bind sites (i.e. those occupied by the substrate) is proportional to its rate, that is $Y \propto v_r$ gave rise to
%\be\label{MMreloaded}
%Y = \frac{k S}{S_0+S},
%\ee
%in proper units: this is one of the most celebrated formulas in Chemistry and it defines the simplest situation where the binding sites do not cooperate, that is, the fact that a site of an enzyme is empty (or full) does not affect in any way the capabilities  of getting empty (or full) of its peers.
%
This model immediately settled down as the paradigm for Chemical Kinetics, somehow similarly to the perfect gas model (where atoms, or molecules - collisions apart - do not interact) of the Kinetic Theory in the early Statistical Physics \cite{gas}. Nevertheless, deviations from this behaviour were not late to arrive: the most common phenomenon was the occurrence of a positive cooperation among the binding sites of a multi-site molecule. Actually, many polymers and proteins exhibit cooperativity, meaning that the ligand binds in a non-independent way: if, upon a ligand binding, the probability of further binding (by other ligands) is enhanced, the system is said to show {\em positive cooperativity}.

\begin{figure}[!ht]
\begin{center}
\includegraphics[width=0.7\textwidth]{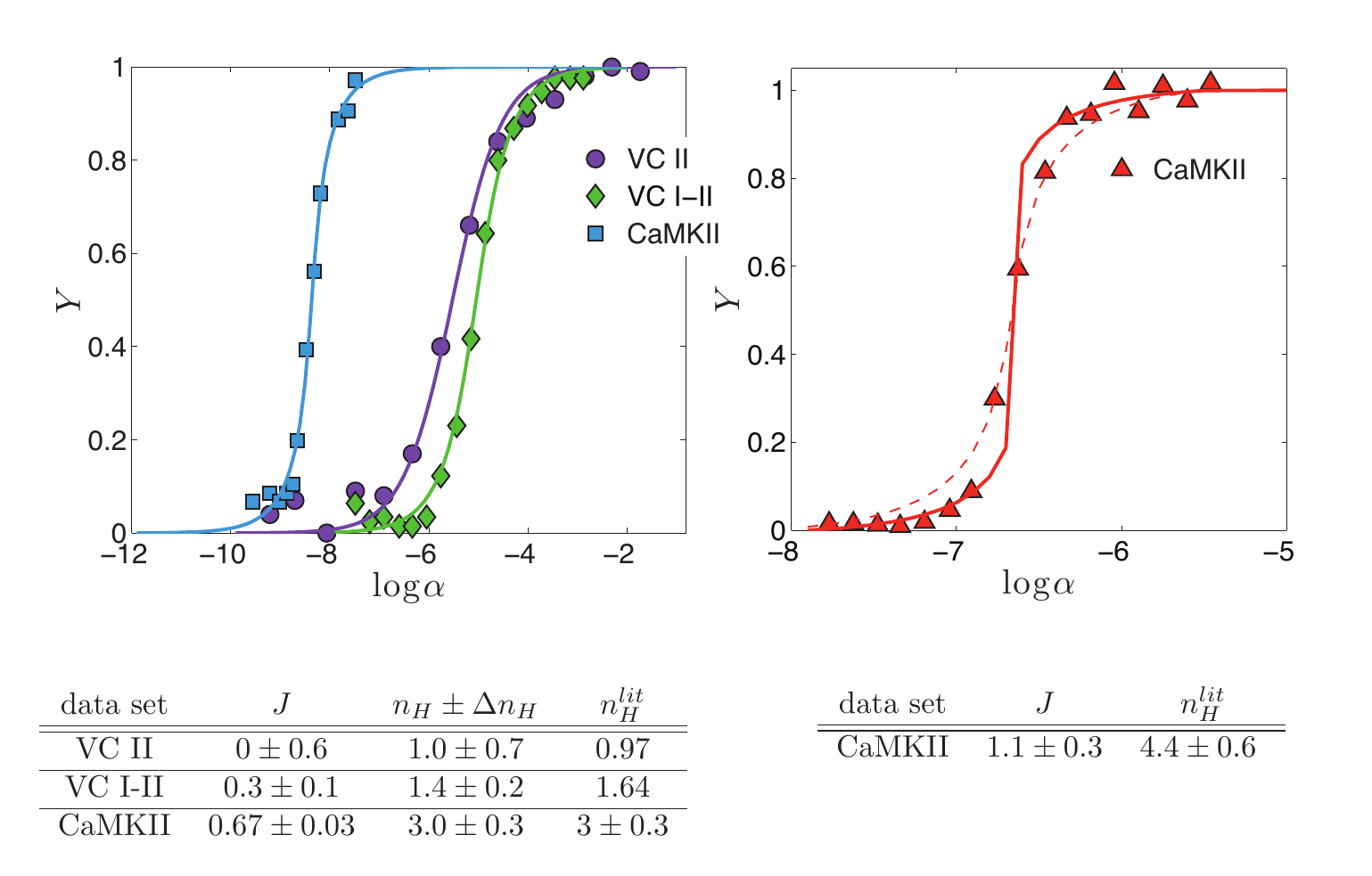}
\caption{These plots show comparison between data from experiments (symbols) and best-fits through Eq.(\ref{selfC}) (lines).
Data refer to non-cooperative and positive-cooperative systems \cite{chao,mandal} (left panel) and an ultra-sensitive system \cite{bradshaw} (right panel). For the latter we report two fits: Dashed line is the result obtained by constraining the system to be cooperative but not-ultra-sensitive (that is, $J \leq 1$), while solid line is the best fit (without constraints) which yields to $J \sim 1.1$, hence a ``first order phase transition" in the language of statistical mechanics. The relative goodness of the fits are $R^2_{coop}\sim 0.85$ and $R^2_{ultra} \sim 0.94$, confirming an ultra-sensitive behavior.
The tables in the bottom present the value of $J$ derived from the best fit and the resulting $n_H$; the estimate of the Hill coefficient taken from the literature is also shown for comparison. This figure was presented in \cite{SciRepBurioni}.}\label{fig:coop}
\end{center}
\end{figure}

To fix ideas, let us make a practical example and let us consider the case of a well-known protein, i.e., the hemoglobin. This is responsible for oxygen transport throughout the body and it ultimately allows cellular respiration. Such features stem from hemoglobin's ability to bind (and to dislodge as needed) up to four molecules of oxygen in a non-independent way: if one of the four
sites has captured an oxygen molecule, then the probability that the remaining three sites will capture further oxygen increases, and vice versa. As a result, if the protein is in an environment rich of oxygen (e.g., in the lungs), it readily binds up to four molecules of oxygen, and, as much readily, it gets rid of them when crossing an oxygen-deficient environment.
To study quantitatively its behaviour one typically measures its characteristic input-output relation. This can be achieved by considering a set of $M$ elementary experiments where these proteins, in the same amount for each experiment, are prepared in a baker and allowed to bind oxygen, which is supplied at different concentrations $S_i$ for different experiments (e.g., $S_1<S_2<...<S_M$). We can then construct a Cartesian plane, where on the abscissas we set the concentration of the ligand $S$ (oxygen in this case, i.e. the {\em input}) while on the $y$-axes we put the fraction of protein bound sites $Y$ (the saturation function, i.e., the {\em output}). In this way, for each experiment, once reached the chemical equilibrium, we get a saturation level and we can draw a point in the considered Cartesian plane; interpolating between all the points  a sigmoidal curve will emerge (see Fig.~\ref{fig:coop}). Archibald V. Hill formulated a description for the behavior of $Y$ with respect to $S$: the so-called Hill equation empirically describes the fraction of molecules binding sites, occupied by the ligand, as a function of the ligand concentration \cite{SciRepBurioni,Aldo,Cattoni,weiss}. This equation generalizes the Michaelis-Menten law (\ref{eq:MM}) and reads as
\be\label{MMreloaded}
Y = \frac{k S^{n_H}}{S_0+S^{n_H}},
\ee
where $n_H$ is referred to as Hill coefficient and can be interpreted as the effective number of binding sites that are interacting each other.
This number can be measured as the slope of the curve $\log[Y/(1-Y)]$ versus $\log(S)$, calculated at the half-saturation point.
%
%The equation is useful for determining the degree of cooperativity of the ligand(s) binding to the enzyme or receptor. The Hill coefficient provides a way %to quantify the degree of interaction between ligand binding sites
%
%
%, whose slope, measured at half-saturation defines the so-called {\em Hill coefficient} $n_H$. The Hill coefficient provides a rough estimate of how cooperative the observed reaction was (how many binding sites are in {\em effective interactions}) and the Hill equation generalizes the Micaelis-Menten one  and reads as
%\be\label{MMreloaded}
%Y = \frac{k S^{n_H}}{S_0+S^{n_H}},
%\ee
Of course, if $n_H = 1$ there is no cooperation at all and each binding site acts independently of the
others (and, consistently, Michaelis-Menten kinetics is restored), viceversa, if $n_H > 1$, the reaction is said to be cooperative (just like in hemoglobin), and if $n_H \gg 1$ the cooperation among binding sites is so strong that the sigmoid becomes close to a step function and the kinetics is named ultra-sensitive.

The Michaelis-Menten law together with the extension by Hill, provided a good description for a bulk of chemical reactions, however, things were not perfect yet. For instance, some yeast's proteins (e.g., the Glyceraldehyde 3-Phosphate Dehydrogenase \cite{Koshland1}) produced novel (mild) deviations from the Hill curve: for these enzymes,  the cooperativity of their binding sites decreases while increasing the ligand concentration.
The following work by Daniel E. Koshland allowed understanding this kind of phenomenology by further enlarging the theoretical framework through the introduction of the concept of negative cooperativity.
In fact, in the previous example, beyond the positive cooperation between the binding sites there are also negative-cooperative effects underlying. Their effective action is to diminish the overall binding capabilities of the enzyme and thus to reduce the magnitude of its Hill coefficient.

\subsection{The Mechanics background}
The progressive enlargement of a theoretical scaffold to fit the always increasing amount of evidences is a common feature in the historical development of scientific disciplines \cite{storia1,storia2}. This is the case also for Mechanics and, as we will see, the analogy with Chemical Kinetics goes far beyond this feature.
\newline
Beyond Kinematics, which describes the motion of systems without considering their mass or the forces that caused the motion, in the seventeen-th century Newton gave a sharp description of Mechanics, in the form of laws describing how masses dynamically respond when stimulated by an external force (or moment). Here, the input is the force while the output is the motion of the body. The Newtonian dynamics has been ruling for centuries and, in fact, it was so well-consolidated that scientists, among which Giuseppe L. Lagrange, William R. Hamilton and Carl G.J. Jacobi, later reformulated the entire theory in a powerful and elegant variational flavor. The theory was overall brilliant to explain the perceivable reality, but with exceptions emerging in the limit of {\em too little} or {\em too fast}.
\newline
We will focus on the latter. In the Newtonian world, if an applied force is kept constant over a mass, this will constantly accelerate, eventually reaching diverging velocities. This was perfectly consistent with the general credo that the speed of light were infinite. However, this postulate broke down in 1887 when the famous experiment by Albert A. Michelson and Edward Morley proved that such a velocity is actually finite. The next years were dense of novel approaches and ideas by many scientists, as Hendrik Lorentz and Hermann Minkowski, and culminated with the special relativity by Albert Einstein in 1905. According to this theory, no mass can exist whose velocity may diverge, the limiting speed being the speed of light. The classical Hamilton-Jacobi equations and Galilean transformations left the place to the Klein-Gordon formulation and Lorentz covariances and contravariances (the natural metric being Minkowskian) \cite{cabibbo}. Clearly, classical mechanics was still a good reference framework for the vast majority of the data collected (much like the positive cooperativity accounted for the bulk of the empirical data in the chemical counterpart), however there were rare phenomena (e.g., a muon decay in atmosphere \cite{majani}) that required a broader scaffold which, in the opportune limits, could recover the classical one.
%
%
%as in the growth of the models in chemical kinetics -where already the positive cooperativity accounted for the bulk of the empirical data, but there were rare oddities as the Koshland enzymes (escaping that framework due to anti-cooperative effects), also , however there were rare phenomena (e.g., a muon decay in atmosphere) that quest for a broader scaffold where all these phenomena could coherently co-exist and that, in the opportune limits, would recover all the previous schemes.

Despite this historical connection between Chemical Kinetics and Classical Mechanics may look weird at a first glance, as we will prove, there is a formal analogy between their mathematical representations. In the next section we will summarize the main results concerning the analogy at the classical level. More sharply, the saturation plot of classical (positive cooperative) chemical kinetics (namely the input-output relation between the saturation function and the concentration of the substrate) can be derived by a minimum action principle that is the same that holds in classical mechanics, when describing a mass motion in the Hamilton-Jacobi framework. In this parallelism, the saturation function in Chemistry plays as the velocity in Physics: thus, exactly as happens in special relativity, the velocity of the mass is bounded (by definition, the saturation function can not exceed one). Indeed, we can follow this mathematical equivalence and verify that there is actually a natural broader formulation for chemical kinetics that is exactly through the Klein-Gordon setting (rather than its classical Hamilton-Jacobi counterpart) and the theory as a whole is Lorentz-invariant. Remarkably, when read with chemical glasses, this extended relativistic setting allows for the anti-cooperative corrections that Koshland revealed in the study of the yeast enzymes, resulting in a complex mixture of positive and negative cooperation among binding sites.

\section{The {\em standard } mathematical scaffold for classical cooperativity}
As anticipated in Sec.~\ref{sec:intro}, cooperativity is a widespread phenomenon in Chemistry and its underlying mechanisms can be multiple: for example, if the adjacent binding sites of a protein can accommodate charged ions, the attraction/repulsion between the ions themselves may result in a positive/negative kinetics; in most common cases, the bonds with the substrate modify the protein conformational structure, by influencing possible further links in an allosteric way \cite{SciRepKatz,Michel}. Whatever the origin, cooperativity in Chemistry is a typical {\em emergent} property that directly relates the microscopic description of a system at the single binding-site level, with the macroscopic properties shown by its constituent molecules, cells, and organisms, thus the use of Statistical Physics for its investigation appears quite natural \cite{SciRepBurioni,Thompson}. Usually, in Statistical Physics one is provided with an (inverse) temperature $\beta$, and with an Hamiltonian (i.e., a cost function) $H(\sigma,J,h)$ describing the model at the microscopic level, namely in terms of elementary variables $\sigma_i, \sigma_j$, couplings among elementary variables $J_{ij}$, and external fields $h_i$ acting over these. The goal is to obtain the {\em free energy} $A(\beta, J, h)$ of the model, from which the average value of the macroscopic observables can be directly derived \cite{Thompson}.

\subsection{Formulation of the problem: the thermodynamical free energy}
In the following we summarize the minimal assumptions needed when modelling chemical kinetics from the Statistical Physics perspective; for a more extensive treatment of this kind of modelling we refer to \cite{Thompson,Aldo,SciRepBurioni,SciRepMoro,SciRepKatz}, while for a rigorous explanation of the underlying equivalence between Statistical Mechanics and Analytical Mechanics we refer to the seminal works by Francesco Guerra \cite{Guerra0}, dealing with the Sherrington-Kirkpatrick model (and then deepened in e.g., \cite{Barra1,Barra4,Barra5,Barra2}), and by Jordan G. Brankov and Valentin Zagrebnov in \cite{Zagrebnov}, dealing with the Husimi-Temperley model (and then deepened in e.g., \cite{Tonio1,Tonio2,Barra-Moro,Giulio}).

\begin{itemize}
\item Each binding site may or may not be occupied by a ligand: this allows us to code its state (empty versus full) by a Boolean variable. For the generic $i^{th}$ site, we will use an Ising spin  $\sigma_i = \pm 1$, where $\sigma_i=-1$ represents an empty $i^{th}$ site, viceversa $\sigma_i=+1$ means that the $i^{th}$ site is occupied. Clearly, if there are overall $N$ binding sites, $i \in (1,...,N)$.

\item It is rather inconvenient, an ultimately unnecessary, to deal with the whole set $\sigma_i,...,\sigma_N$ if we are interested in the properties of large numbers of these variables (i.e., in the so called {\em thermodynamic limit} corresponding to $N \to \infty$). If we want to distinguish between a fully empty state $\sigma_i=-1 \forall i \in (1,...,N)$ (ordered case), a fully occupied state $\sigma_i=+1 \forall i \in (1,...,N)$ (ordered case) or a completely random case where $\sigma_i = \pm 1$ with equal probability (disordered case), it is convenient to introduce the {\em order parameter} for these variables as the {\em magnetization} $M$ (this term stems from the original application of the Statistical Mechanics model in the context of magnetism) that reads as the arithmetic average of the spin state, namely
\be
M = \frac1N \sum_{i=1}^N \sigma_i \in [-1,+1].
\ee
There is a univocal relation between the magnetization in Physics and the saturation function in Chemistry, where, we recall, we denote with $Y \in [0,1]$ the fractional occupation of the binding sites. In fact, one has \cite{Aldo,SciRepBurioni}
\be\label{ponte1}
Y = \frac12 \left(1 + M\right).
\ee
{\em Eq.(\ref{ponte1}) constitutes the first bridge between the Chemistry we aim to describe (via the saturation function $Y$) and the Physics that we want to use (via the magnetization $M$)}.

\item All the binding sites interact with the ligand by the same strength. This is a standard assumption in Chemical Kinetics \cite{Aldo,Ricci1,weiss} and it means that the diffusion of the ligands is fast enough to ensure a homogeneous solution. The concentration of free ligands is mapped into a one-body contribution $H_1$ in the cost-function. This term encodes for the action of an \emph{external magnetic field} in such a way that, if the field acting on the $i^{th}$ is positive, the spin will tend to align upwards (namely this direction is energetically favored), and vice versa.
This homogenous mixing assumption translates into a homogeneous external field $h$, and the related contribution reads as
\be\label{corpouno}
H_1(\sigma,h) = - h \sum_{i=1}^N \sigma_i.
\ee
Notice that $h$ plays as a chemical potential and, consistently, it can be related to the substrate concentration as
\be\label{ponte2}
h = \frac12 \log \left(\frac{S}{S_0} \right),
\ee
$S_0$ being the value of the ligand concentration at half saturation.
\newline
{\em Eq.(\ref{ponte2}) constitutes the second bridge between the Chemistry we aim to describe (via the ligand concentration $S$) and the Physics that we want to use (via the magnetic field $h$)}.

\item The binding sites can cooperate in a positive manner: this can be modelled by introducing a coupling between the $\sigma$ variables. The simplest mathematical form is given by a two-body contribution $H_2$ in the cost function. This term encodes for the reciprocal interactions among binding sites and it reads as
\be
H_2(\sigma,J) = - \frac{J}{N} \sum_{i<j}^N \sigma_i \sigma_j,
\ee
where $J \geq 0$ is the interaction strength and the sum runs over all possible pairs; the normalization factor $1/N$ ensures the linear extensivity of the cost-function with respect to the system size. A positive value for $J$ implies an imitative interaction among binding sites: configurations where spins tend to be aligned each others (namely where sites tend to be either all occupied or all unoccupied) are energetically more favoured and will therefore be more likely.

\item Combining together the previous contributions we get the total Hamiltonian
\be\label{CiVu}
H(\sigma,J,h)=H_1(\sigma,h)+H_2(\sigma,J)=-\frac{J}{N}\sum_{i<j}^N \sigma_i \sigma_j - h \sum_{i=1}^N \sigma_i= -N \left( \frac{J}{2N}M^2 + h M \right).
\ee
It is possible to introduce the free energy associated to such an Hamiltonian as
\begin{eqnarray}\label{laF}
A(\beta,J,h) &=& \frac{1}{N}\log \sum_{\sigma_1,...,\sigma_N}^{2^N} \exp\left[ -\beta H(\sigma,J,h) \right]\\
&=& \frac{1}{N}\log \sum_{\sigma_1,...,\sigma_N}^{2^N} \exp\left( \frac{\beta J}{2N}\sum_{i,j}^N \sigma_i \sigma_j + \beta h \sum_i^N \sigma_i \right),
\end{eqnarray}
where $\beta$ is the inverse temperature in proper units.
The free energy is a key observable because it corresponds to the difference between the internal energy $U$ and the entropy $S$ (at given temperature), i.e. $A(\beta,J,h) = S(\beta,J,h) - \beta U(\beta,J,h)$. If we could obtain an explicit expression for $A(\beta,J,h)$ in terms of the order parameter $M$, we could obtain an expression for the magnetization expected at equilibrium by imposing $\delta_M A(\beta,J,h)=0$, in fact, this implies that we are simultaneously asking for the minimum energy and the maximum entropy.% (note that we extremize the free energy functional w.r.t. $M$ keeping the free variables fixed).
%
%In other words, the solutions that we could find by this procedure would be all and solely the thermodynamical solutions, that is what we actually want.
%\newline

Notice that, having stated the two bridges given by Eq.s(\ref{ponte1}) and (\ref{ponte2}), other mappings between the two fields (e.g., the relation between the coupling strength $J$ and the Hill coefficient $n_H$, see eq.(\ref{HillDef}) later on) emerge spontaneously as properties of the thermodynamic solutions of the problems.
\end{itemize}

\subsection{Resolution of the problem: the mechanical action}
We want to find an explicit expression (in terms of $M$) for the free energy defined in eq. (\ref{laF}). To this task let us rename $-\beta J = t$ and $\beta h = x$ and let us think of these fictitious variables as a time and a space, respectively. Thus, we can write the free energy as
\be\label{action}
A(t,x) = \frac{1}{N} \log \sum_{\sigma_1,...,\sigma_N}^{2^N} \exp\left( \frac{-t}{2N}\sum_{i,j}^N \sigma_i \sigma_j + x \sum_i^N \sigma_i \right),
\ee
where we also wrote $\sum_{i<j}\sigma_i \sigma_j \sim (1/2) \sum_{i,j} \sigma_i \sigma_j$, which implies vanishing corrections in the thermodynamic limit.
\newline
If we work out the {\em spatial} and {\em temporal} derivatives of the free energy (\ref{action}) we obtain
\begin{eqnarray}\label{stream1}
\frac{d A(t,x)}{dt} &=& -\frac{1}{2} \langle M^2 \rangle_{t,x},\\  \label{stream2}
\frac{d A(t,x)}{dx} &=&  \langle M \rangle_{t,x},
\end{eqnarray}
where the average $\langle \cdot \rangle_{t,x}$ for a generic observable $O$ depending on the spin configuration
%$\sigma_1, \sigma_2, ..., \sigma_N$
is defined as
\begin{equation}
\braket{O}_{t,x}=\dfrac{\sum_{\sigma} O \exp \left[N (-t \cdot M^2 + x \cdot M)\right]}{\sum_{\sigma}  \exp\left[ N (-t \cdot M^2 + x \cdot M)\right]},
\end{equation}
and, posing $t=-\beta J$ and $x=\beta h$, the Boltzmann average for the original system (\ref{CiVu}) is recovered and this shall be simply denoted as $\langle \cdot \rangle$
%
%
%Boltzmann averages $\braket{\dots}$ are defined as (using the magnetization as a trial function)
%\begin{equation}
%\braket{M}_{t,x}=\dfrac{\sum_{\sigma} M \exp \left(N (-t \cdot M^2 + x \cdot M)\right)}{\sum_{\sigma}  \exp\left(N (-t \cdot M^2 + x \cdot M)\right)}.
%\end{equation}
%The averages $\braket{\dots}_{t,x}$ will be denoted by $\braket{\dots}$ whenever evaluated in the sense of Thermodynamics (i.e. for $t=-\beta J$ and $x=\beta h$).
\newline
If we now introduce a potential $V(t,x)$, defined as half the variance of the magnetization, i.e.,
\be
V(t,x)= \frac12 \left( \langle M^2 \rangle - \langle M \rangle^2 \right),
\ee
we see that, by construction, the free energy of this model obeys the following Hamilton-Jacobi equation
\be\label{HJ1}
\frac{d A(t,x)}{dt} + \frac12 \left( \frac{d A(t,x)}{dx} \right)^2 + V(t,x)=0,
\ee
and therefore $A(t,x)$ is also an {\em action} of Classical Mechanics.
We can simplify the previous equation by noticing that, for large enough volumes, the magnetization is a self-averaging quantity \cite{Barra1,Thompson}, thus in the infinite volume limit the potential must vanish, that is, $\lim_{N \to \infty} V(t,x) = 0$.
Here, we are restricting to large volumes and we are therefore left with a Hamilton-Jacobi equation describing a free propagation; since the potential is zero, the Lagrangian $\mathcal{L}$ coupled to the motion is just the kinetic term
\be\label{Lagrange}
\mathcal{L}=\frac12 \langle M^2 \rangle,
\ee
that is the analogous of the classical formula $\mathcal{L} = \frac12 m v^2$ where the mass $m$ is set unitary (i.e., $m =1$), and the role of the velocity $v$ is played by the average magnetization $\langle M \rangle$.
Solving the Hamilton-Jacobi equation is then straightforward: the solution is formally written as
\be
A(t,x)=A(t=0,x=x_0) + \int_0^t \mathcal{L}(t',x) dt'.
\ee
The evaluation of the Cauchy condition $A(t=0,x=x_0)$ is trivial because, at $t=0$, the coupling between variables disappears (see eq.(\ref{laF})), while the integral of the Lagrangian over time reduces to the Lagrangian times time (as the potential is zero).
Pasting these two contributions together we obtain
\be
A(t,x)= \ln 2 + \ln\cosh\left( x_0 \right) + \frac12 \langle M^2 \rangle \cdot t.
\ee
Finally, noticing that the equation of motion is a Galilean trajectory as $x(t)=x_0 + \langle M \rangle  t$ (hence $x_0=x - \langle M \rangle t$) and recasting the solution back in the original variables, i.e. $t=-\beta J$ and $x=\beta h$, we get the free energy associated to this general positive cooperative reaction:
\be\label{FECK}
A(\beta,J,h)= \ln 2 + \ln\cosh\left( \beta J \langle M \rangle + \beta h \right) - \frac12 \beta J \langle M^2 \rangle.
\ee
By extremizing $A(\beta,J,h)$ with respect to $\langle M \rangle$ we get
\be\label{self1}
\frac{d A(\beta,J,h)}{d\langle M \rangle}=0 \Rightarrow \langle M \rangle = \tanh\left[\beta \left(J \langle M \rangle + h \right)  \right].
\ee
This result recovers the well known self-consistency equation for the order parameter of the Curie-Weiss model in Statistical Mechanics \cite{Barra1,Thompson}.

\subsection{Chemical properties of the physical solution} \label{sec:prop}
The self-consistent equation in Eq.(\ref{self1}) is an input-output relation for a general system of binary elements, possibly positively interacting, under the influence of an external field: the input in the system is the external field $h$ and the output is the magnetization $M$. We can rewrite Eq.(\ref{self1}) in a chemical jargon by using the bridges coded in the Eq.s (\ref{ponte1}) and (\ref{ponte2}) and fixing, for the sake of simplicity, $S_0=1$, that is
\be\label{selfC}
Y(J,S)=\frac12 \left[1 + \tanh\left( J(2 Y -1) + \frac12 \ln S \right)\right] = \frac{S e^{2J (2Y-1)}}{1+S e^{2J (2Y-1)}}.
\ee
Before proceeding, we check that if cooperation disappears (i.e., binding sites are reciprocally independent), the Michaelis-Menten scenario is recovered. Posing $J=0$ in the equation above we get
\be\label{MSMM}
Y(J=0,S) = \frac{S}{1+S},
\ee
that is (apart a constant factor that can be re-introduced by taking $S_0=k$, rather than $1$) the Michaelis-Menten equation (see Eq.~\ref{eq:MM}).

One step forward, we now take into account the coupling $J$ and relate it to the Hill coefficient $n_H$. The latter is defined in Chemistry as the slope of $Y(S)$ at half saturation (i.e., when $Y=1/2$), and we can obtain its expression following this prescription by using Eq.(\ref{selfC}), namely
\be\label{HillDef}
n_H = \frac{1}{Y(1-Y)} \left. \frac{d Y}{ d S} \right |_{Y=\frac{1}{2}} = \frac{1}{1-J}.
\ee
We note that as $J \to 0$ we get, as expected, $n_H \to 1$: if there is no cooperation between binding sites, the Hill coefficient must be unitary; further, the stronger the coupling $J$ and the (hyperbolically) larger the value of the Hill coefficient. In particular, for $J \rightarrow 1$ the kinetics gets ultra-sensitive and discontinuities emerge. We remark that, with simpler statistical mechanics model as linear chains of spins, phase transitions are not allowed, hence ultra-sensitive behavior can not be taken into account: the present framework is the simplest non-trivial scheme where all these phenomena can be recovered at once (see Fig.~\ref{fig:coop} and \cite{SciRepBurioni} for more details on ultra-sensitive kinetics).
\newline
Also, it is worth highlighting the full consistency between our treatment of ultra-sensitive kinetics and more standard ones as for instance reported in \cite{bookMazza} (see eq.$5.17$ therein), where the expression for the Hill coefficient can be translated into our formulation as
$$
n_H = \frac{N \left( \langle M^2 \rangle - \langle M \rangle^2 \right)}{1-\langle M \rangle^2}.
$$
One we see that for $\langle M \rangle \to \pm 1$ the Hill coefficient diverges, which is the signature of an ultra-sensitive behavior: this is perfectly coherent with our approach where, in that limit, the input-output relation (see the hyperbolic tangent (\ref{selfC})) becomes a step function.

However, as mentioned in the Introduction, this theory has its flaws, in Chemistry as well as in Mechanics. Regarding the former, the complex picture of yeast's enzymes evidenced by Koshland \cite{Koshland1,Koshland2}, where positive and negative cooperativity appear simultaneously (and with the anti-cooperativity effect getting more and more pronounced as the substrate concentration is raised), still escapes from this mathematical architecture. Further, from the mechanical point of view, two weird things happen: the velocity $M$ is bounded by $c=1$, while in Classical Mechanics the velocity may diverge; further, if we look at the Boltzmannfaktor in the free energy (Eq.~\ref{action}), this reads as $\exp \left[N \left(  -t M^2/2 + x M \right)\right]$ and, recalling that the kinetic energy in this mechanical analogy reads as $M^2/2$ (the mass is unitary, thus velocity and momentum coincide), we are allowed to interpret $A(\beta,J,h)$ as a real action. From this perspective, the exponent can be thought of as the coupling between the stress-energy tensor and the metric tensor: a glance at the form of the Boltzmannfaktor reveals that the natural underlying metric is $(-1,+1)$ rather than $(+1,+1)$ as in classical Euclidean frames, or in other words, it is of the Minkowskian type. All these details point toward the generalization of the equivalence including special relativity.
\newline
Plan of the next section is to follow the mechanical path and extend the classical kinetic energy including relativistic corrections and then to investigate its implications. We will see that in the broader, relativistic framework for chemical kinetics the deviations that Koshland explained adding an anti-cooperative interactions -beyond the cooperative ones- at high ligand's doses are the chemical analogies of the deviation from classical mechanics at high velocities observed in special relativity.

\section{The {\em generalized} mathematical scaffold for mixed cooperativity}
\subsection{Relativistic setting}
The relativistic extension of the the Hamiltonian (\ref{CiVu}) is defined by an Hamiltonian of the form
\begin{equation}
\label{eq:rel}
\dfrac{H(\sigma,J,h)}{N}=-J\sqrt{1+M^2} - h M,
\end{equation}
 where $M=\dfrac{1}{N}\sum_i^N\sigma_i$ as usual. Next, we have to insert (\ref{eq:rel}) into the free energy (\ref{laF}):
 \begin{equation}
 \label{eq:freeR}
 A(t,x)=\dfrac{1}{N} \log \sum_{\sigma}^{2^N} \exp (-t \cdot N  \sqrt{1+M^2} + x \cdot M),
 \end{equation}
where we already replaced $t=-\beta J$ and $x=\beta h$ in order to work out their streaming, that read as
\begin{equation}
\begin{split}
& \dfrac{\partial A(t,x)}{\partial t}=-\braket{\sqrt{1+M^2}}_{t,x}, \\
& \dfrac{\partial A(t,x)}{\partial x}=\braket{M}_{t,x}, \\
& \dfrac{\partial^2_{tt}A(t,x)-\partial^2_{xx} A(t,x)}{N}=1-\braket{\sqrt{1+M^2}}_{t,x}^2+\braket{M}_{t,x}^2, \\
\end{split}
\end{equation}
where the Boltzmann averages $\braket{\dots}_{t,x}$ are defined as (using the magnetization as a trial function)
\begin{equation}
\label{eq:Boltz}
\braket{M}_{t,x}=\dfrac{\sum_{\sigma} M \exp \left[N (-t  \cdot \sqrt{1+M^2} + x \cdot M)\right]}{\sum_{\sigma}  \exp\left[N (-t \cdot  \sqrt{1+M^2} + x \cdot M]\right)}.
\end{equation}
As before, the averages $\braket{\dots}_{t,x}$ will be denoted by $\braket{\dots}$ whenever evaluated in the sense of thermodynamics (i.e. for $t=-\beta J$ and $x=\beta h$).
By a direct calculation, it is straightforward to see that the expression (\ref{eq:freeR}) obeys the relativistic Hamilton-Jacobi equation
 \begin{equation}\label{eq:HJR}
 \begin{split}
&(\partial_t A)^2 -(\partial_x A)^2 + V_N(t,x)=1, \\
& V(t,x)=\dfrac{1}{N}\Box A(t,x),   \\
\end{split}
 \end{equation}
where the symbol $\square$ represents the D'Alambert operator and $V(t,x)$ is the potential whose expression is chosen in order to make the equation valid by construction and, this time, it is automatically Lorentz invariant. If the functional $A(t,x)$ is sufficiently smooth (that is, its derivatives are regular functions of $t$ and $x$), in the thermodynamic limit, we have $\lim_{N\rightarrow \infty} V(t,x)=0$, hence in this high-volume limit we are left with
\begin{equation}
(\partial_{\mu} A)^2=1,
\end{equation}
which is the Klein-Gordon equation for a free relativistic particle with unitary mass in natural units ($m_0 =1$). \\
In relativistic mechanics, the stress energy tensor of this particle is defined as
\begin{equation}
\label{eq:momentum1}
p^{\mu}=\Big ( E, \gamma v\Big ),
\end{equation}
where $v$ is the classical velocity of the particle, $\gamma=1/\sqrt{1-v^2}$ and $E=\gamma m_0 =\gamma$ is the relativistic energy. In addition, the contravariant momentum is expressed through the action by the following equation
\begin{equation}
\label{eq:momentum2}
p^{\mu}=-\dfrac{\partial A}{\partial x_{\mu}} = (\braket{\sqrt{1+M^2}}_{t,x},\braket{M}_{t,x}).
\end{equation}
Comparing (\ref{eq:momentum1}) and (\ref{eq:momentum2}), it is immediate to identify the magnetization as the relativistic dynamical variable:
\begin{equation}
\label{eq:magn}
\braket{M}_{t,x}=\dfrac{v}{\sqrt{1-v^2}},
\end{equation}
while the Lorentz factor is
\begin{equation}
\label{eq:Lorentz}
\gamma=\sqrt{1+\braket{M}^2_{t,x}}.
\end{equation}
In the thermodynamic limit, the particle is free and its trajectories are the straight lines $x=x_0+vt$. Since the relativistic Lagrangian $\mathcal{L}=-\gamma^{-1}$ is constant along these classical trajectories, the free energy can be computed as:
\begin{equation}
\label{eq:Lagrangian}
\begin{split}
A(t,x)&=A(0,x_0)+\int_0^t dt' \, \mathcal{L}(t')=A(0,x_0)-\dfrac{t}{\gamma} =\log 2 + \log \cosh (x_0) -\dfrac{t}{\gamma}=\\
&=  \log 2  + \log \cosh (x-vt) - \dfrac{t}{\sqrt{1+\braket{M}^2_{t,x}}}  = \\
& = \log2 + \log \cosh \left( x- \frac{\braket{M}_{t,x} t}{\sqrt{1+\braket{M}_{t,x}^2}} \right)-\dfrac{t}{\sqrt{1+\braket{M}^2_{t,x}}}.  \\
\end{split}
\end{equation}
\noindent Setting $t=-J\beta$ and $x=\beta h$, we finally get an explicit expression for the free energy:
\begin{equation}
\label{eq:free1}
A(\beta)=\log 2 + \log \cosh \left( \frac{J \beta \langle M \rangle}{\sqrt{1+\braket{M}^2}}  \right) +\dfrac{J \beta}{\sqrt{1+\braket{M}^2}}.
\end{equation}
Requiring that the free energy is extremal with respect to the magnetization (that from a thermodynamical perspective can be seen as the simultaneous effect of the minimum energy and the maximum entropy principles, and, from a mechanical perspective as the minimum action principle), the associated self-consistency equation becomes
\begin{equation}
\label{eq:self}
\braket{M}=\tanh \left ( \dfrac{J \beta \braket{M}}{\sqrt{1+\braket{M}^2}} +\beta h \right ).
\end{equation}
%Under the influence of an external field $h$, the relativistic self-consistency relation is
%\begin{equation}
%\label{eq:magnetico}
%\braket{M}=\tanh \left ( \dfrac{J \beta \braket{m}}{\sqrt{1+\braket{m}^2}} + \beta h \right ).
%\end{equation}

\subsection{The classical limit from a chemical perspective} \label{sec:classicallimit}
Reading the self-consistency (\ref{eq:self}) in chemical terms, that is using the bridges Eq.s (\ref{ponte1},\ref{ponte2}), we obtain
%\begin{equation}
%\label{eq:saturation1}
%Y(S,J)=\dfrac{1}{2} \left[ 1+\tanh \left ( \dfrac{J \beta \braket{m}}{\sqrt{1+\braket{m}^2}} + \dfrac{\beta}{2} \log \left( \dfrac{S}{S_0}\right) \right )\right].
%\end{equation}
%Inverting the (\ref{eq:saturation}), the final formula for the saturation is
\begin{equation}
\label{eq:saturation2}
Y(S,J)=\dfrac{1}{2} \left[ 1+\tanh \left ( \dfrac{J \beta (2Y-1)}{\sqrt{1+(2Y-1)^2}} + \dfrac{\beta}{2} \log \left( \dfrac{S}{S_0}\right) \right )\right].
\end{equation}
We can now check whether, under suitable conditions, this broader theory recovers the classical limit.
First, we notice that under the assumption of no interactions among binding sites (i.e., $J=0$) and replacing $h=\frac12 \log (S/S_0)$, the Michaelis-Menten behaviour is recovered. This can be shown by rewriting Eq.~\ref{eq:saturation2} as
\begin{equation}
\label{eq:saturation3}
Y(S,J)=\dfrac{S e^{\frac{2J \beta (2Y-1)}{\sqrt{1+(2Y-1)^2}}}}{1+S e^{\frac{2J \beta (2Y-1)}{\sqrt{1+(2Y-1)^2}}}},
\end{equation}
where we also shifted $S/S_0 \rightarrow S$ for simplicity.
For $J=0$ the previous equation reduces to $Y(S)= S/(1+S)$.
%, and by shifting $S\rightarrow k^{-1}S$ the more familiar Michaelis-Menten expression () is recovered. %Expanding the (\ref{eq:saturation3}) at the first order in $J$, we obtain
%
Further, taking the classical limit, at the lowest order, we have the following expansions
\begin{equation}
\label{eq:expansion2}
\begin{split}
& \dfrac{1}{1+\braket{M}^2}=1-\dfrac{\braket{M}^2}{2} + \mathcal{O} \left( \braket{M}^3 \right),\\
& \dfrac{\braket{M}}{1+\braket{M}^2}=\braket{M} + \mathcal{O} \left( \braket{M}^3 \right), \\
\end{split}
\end{equation}
such that (\ref{eq:self}) reduces to (\ref{self1}), in the physical context, and to (\ref{selfC}), in chemical context.
%the self-consistency condition for the order parameter $M$ reads as
%\begin{equation}
%\label{eq:nonrel}
%\braket{m}=\tanh ( \beta J \braket{m} + \beta h).
%\end{equation}
%Once the classical framework is recovered, the saturation function becomes
%\begin{equation}
%\label{eq:nonrel1}
%Y(S,J)=\dfrac{1}{2} \left[ 1+\tanh \left( J(2Y-1)+\dfrac{1}{2} \log \dfrac{S}{S_0}, \right) \right]
%\end{equation}
Clearly, also the slope at $Y=1/2$ is preserved hence, in the classical limit, we recover the expected expression for Hill coefficient (see Eq.~\ref{HillDef}), namely
\begin{equation}
\label{eq:Hill}
n_H=\dfrac{1}{Y(1-Y)}\dfrac{\partial Y}{\partial S} \Bigg | _{Y=1/2}=\dfrac{1}{1-J}.
\end{equation}

\subsection{Beyond the classical limit}

To understand why we expect variations with respect to the Hill paradigm at relatively large values of the substrate concentration, we must check carefully the relativistic self-consistency (\ref{eq:self}). Let us assume we are working at not-too high velocities (that is $\langle M \rangle < 1$) and we can expand the argument inside the hyperbolic tangent, in particular, approximating $1/(\sqrt{1+x^2}) \sim 1 - x^2/2$ , we get
\be\label{approx}
\langle M \rangle = \tanh\left( \beta J \langle M \rangle  - \frac{\beta J}{2} \langle M^3 \rangle + \beta h \right).
\ee
The relativistic effects in chemical kinetics become transparent in this way: if we look at the field felt by the binding sites (i.e., the argument inside the hyperbolic tangent), we see that, beyond the standard Hill term $\beta J \langle M \rangle$ (that positively pairs binding sites together) another term appears that, this time, negatively pairs binding sites together. Retaining this level of approximation, we could write an effective Hamiltonian to generate Eq. (\ref{approx}) that reads as
\be\label{Happrox}
H(\sigma,J,h)= -\frac{J}{N} \sum_{i<j}^N \sigma_i\sigma_j + \frac{J}{2N^3}\sum_{i<j<k<l}\sigma_i \sigma_j \sigma_k \sigma_k,
\ee
hence, beyond the two-body positive coupling coded by the first term, another four-body negative coupling appears. The latter is responsible for the deviation from the classical paradigm and these deviations are in full agreement with the Koshland generalization toward the concept of mixed positive and negative cooperativity \cite{Koshland1}.
\newline
In particular, we can see at work the entire reasoning of Koshland who pointed out how, at large enough substrate concentration, the positivity of the reaction diminishes. In fact, for $\langle M \rangle \sim 0$ no relativistic effect can be noted. By increasing $S$ (the input in the system), we get a growth in $\langle M \rangle$ (the output in the system): the latter raises in response of $S$ and it is enhanced because of the two-body term in the effective Hamiltonian (\ref{Happrox}), the four-body term still being negligible. As $S$ keeps on growing, $\langle M \rangle$ increases as well, up to a point where it reaches high enough values such that, from now on, also the four-body term inside the effective Hamiltonian (\ref{Happrox}) becomes relevant. At this point, a novel, anti-cooperative effect is naturally induced in the reaction and it yields to a reduction of the Hill coefficient. In the next analysis these qualitative remarks shall be addressed in more details.

We focus on the definition of the Hill coefficient based on the Hill equation
\be \label{eq:plot}
Y = \frac{S^{n_H}}{k + S^{n_H}}.
\ee
This equation accounts for the possibility that not all
receptor sites are independent: here $n_H$ is the average number of interacting sites and the slope of the Hill plot. The latter is based on a linear transformation made by rearranging Eq.~(\ref{eq:plot}) and taking the logarithm:
 \be
 \log \left ( \frac{Y}{1-Y} \right) = n \log(S) - \log(k).
 \ee
Thus, one plots $\log Y/(1 - Y)$ versus $\log S$, fits with a linear function and the resulting slope, calculated at the half-saturation point, provides the Hill coefficient. As already underlined, the Michaelis-Menten theory corresponds to $n_H=1$ and any deviations from a slope of $1$ tell us about deviation from that model.

For the (classical and relativistic) models analyzed here (coded in the Hamiltonians (\ref{CiVu}) and (\ref{eq:rel})) we can estimate the slope $n_H$ directly from the self-consistency equations (\ref{selfC}) and (\ref{eq:saturation2}).
Let us start with the classical model. We preliminary notice that
\be \label{eq:HillP}
\frac{d}{d \log(S)} \left. \log \left ( \frac{Y }{1 - Y}  \right) \right \rvert_{Y=1/2} = \frac{1}{Y(1-Y)} \left. \frac{dY}{d \log (S)} \right \rvert_{Y=1/2} = 4 \left. \frac{dY}{d \log (S)} \right \rvert_{Y=1/2}.
\ee
Therefore, we just need to evaluate $dY / d \log(S)$ in $Y=1/2$, which reads as
\be
\left. \frac{d Y}{d \log(S)} \right \rvert_{Y=1/2} = \left. \frac{1}{2} \textrm{sech}^2 \left [J(2Y-1) + \frac{1}{2} \log(S) \right] \left( 2 J \frac{dY}{d \log(S)} + \frac{1}{2} \right) \right \rvert_{Y=1/2}.
\ee
Posing $x = dY / d \log(S) |_{Y=1/2}$ and noticing that $S=1$ when $Y=1/2$, we have
\be
x = \frac{1}{2} \left(2 J x + \frac{1}{2} \right) \Rightarrow x = \frac{1}{4} \frac{1}{1-J}.
%&=& \frac{1}{2} \left ( \frac{2}{ \sqrt{S}} + \frac{1}{\sqrt{S}} \right)^2 \left(2 J x + \frac{1}{2} \right)
\ee
By plugging this result in Eq.~\ref{eq:HillP}, we finally have
\be
n_H^{class} =  \frac{1}{1-J}.
\ee
One can see that when $J=0$ the Hill coefficient is unitary as expected for non-cooperative systems, when $J>0$ the coefficient is larger than 1, indicating that receptors are interacting, and when $J<0$ the coefficient is smaller than 1, as expected for negative cooperativity.

Let us now move to the relativistic model. Again, we just need to evaluate $dY / d \log(S)$ in $Y=1/2$, which, recalling (\ref{eq:saturation2}), reads as
\begin{eqnarray}
\left. \frac{d Y}{d \log(S)} \right \rvert_{Y=1/2} &=&  \frac{1}{2} \textrm{sech}^2 \left [ \frac{J(2Y-1)}{\sqrt{1 + (2Y-1)^2}} + \frac{1}{2} \log(S) \right] \\
&\times&\left. \left(\frac{2J}{[2 + 4 Y (Y-1)]^{3/2}} \frac{dY}{d \log(S)} + \frac{1}{2} \right) \right \rvert_{Y=1/2}.
\end{eqnarray}
Exploiting the fact that $S=1$ when $Y=1/2$, the previous expression simplifies as
\be
\left. \frac{d Y}{d \log(S)} \right \rvert_{Y=1/2} \frac{1}{4} \frac{1}{1-\frac{J}{\sqrt{27}}}.
\ee
Thus, we can write
\be \label{eq:HillPrel}
n_H^{rel} = \frac{d}{d \log(S)} \left. \log \left ( \frac{Y }{1 - Y}  \right) \right \rvert_{Y=1/2} = 4 \left. \frac{dY}{d \log (S)} \right \rvert_{Y=1/2} =  \frac{1}{1-\frac{J}{\sqrt{27}}}.
\ee
Note that $n_H^{class}/n_H^{rel} < 1$, confirming that the relativistic correction weakens the emerging cooperativity.
%In fact, in this Hill-plot framework, the effective coupling for the relativistic model is reduced by an average factor $\approx 0.19$.

\begin{figure}
    \includegraphics[width=0.7\textwidth]{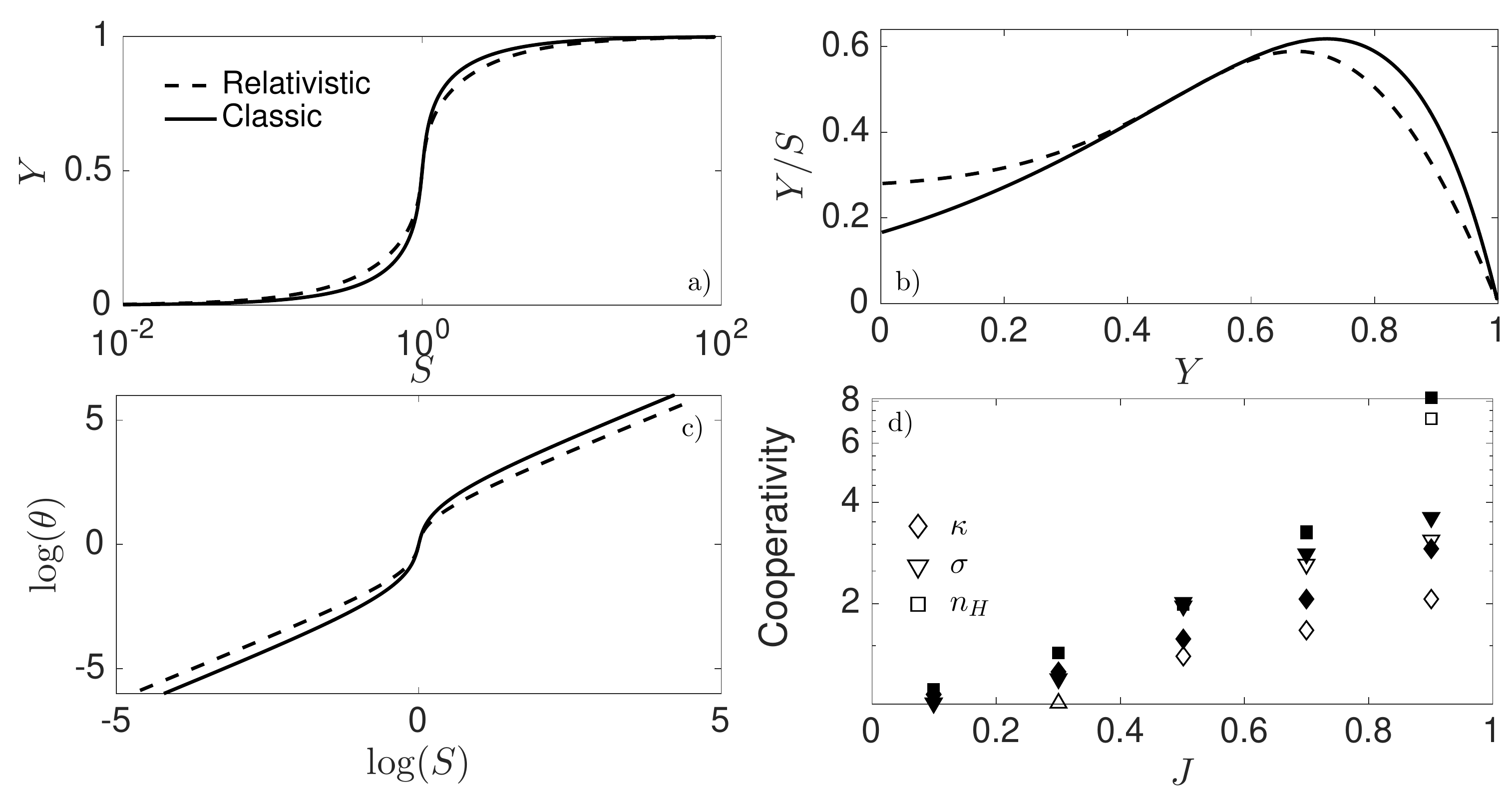}
    \caption{\textbf{Summary of analysis on cooperativity}. Panel a) Klotz plot showing the saturation function $Y$ versus the substrate concentration $S$ (notice the logarithmic scale on the $x$-axis) for the relativistic (dashed line) and the classical (solid line) models. Data for $Y$ are collected by solving numerically the self-consistent equations (Eqs.~(\ref{eq:saturation2}) and (\ref{selfC}), respectively) for $J=0.9$ and different values of $S$. Both models exhibit the sigmoidal shape typical of cooperative systems, however, the former displays a slower saturation. Analogous results are obtained for different values of $J > 0$.  Panel b)  A Scatchard plot is built with the same collection of data by showing the ratio $Y/S$ versus $Y$. Both models exhibit the concave-down shape typical of cooperative systems. However, for relatively small values of $J$ the relativistic model is monotonically decreasing (see also Fig.~\ref{fig:due}). Panel c) A Hill plot is built with the same collection of data by showing $\theta = Y/(1-Y)$ versus $S$; both observables are taken under the logarithm. When $S$ is close to one (here $S_0=1$) the relativistic and the classical model give overlapped curves, while when $S$ is either very large or very small the two curves are shifted. Panel d) By further analyzing the plots in the previous panels we can derive estimates for the extent of cooperativity characterizing the systems. As explained in the main text, starting from data in panel a) we measured the Kloshand quantifier $\kappa = S_{0.9}/S_{0.10}$ ($\diamond$), by extrapolating the maximum value for data in panel b) we get $\sigma$ ($\triangledown$), and by fitting the data in panel c) at the half-saturation point we get $n_H$ ($\square$). These estimates are obtained for both the relativistic (white symbol) and for the classic (black symbols) models.
    %The estimate stemming from the statistical mechanics model $1/(1-J)$ is as well reported ($\circ$).
    }
            \label{fig:uno}
\end{figure}

\subsection{Further robustness checks}
%
%We now introduce further (standard) positivity quantifiers in Chemistry to compare the outcomes of how the two theories, relativistic and its classical limit, perform with these other parameters.
%\newline
As stressed above, for a fixed interaction coupling $J$, the relativistic model is expected to exhibit a lower cooperativity with respect to the classical model. In order to quantify this point we considered different quantifiers for cooperativity and we compared the outcomes for the relativistic and the classical models set at the same value of $J$.
Let us start with the Koshland measure of cooperativity which is defined as the ratio\footnote{Notice that the Koshland index $\kappa$ is actually strongly related to the Hill coefficient (see e.g., \cite{bookMazza}).}
\be
\kappa = \frac{R_{0.9}}{R_{0.1}},
\ee
where $R_{0.9}$ denotes the substrate concentration corresponding to a $90 \%$ saturation, while $R_{0.1}$ denotes the substrate concentration corresponding to a $10 \%$ saturation, that is, $Y(R_{0.9}) =0.9$ and $Y(R_{0.1}) =0.1$. In the non-cooperative case one has $R_{0.9}/R_{0.1} = 81$ and, accordingly, if the ratio is smaller than $81$ (meaning that the saturation curve is relatively steep) one has positive cooperativity, while if the ratio is larger than $81$ one has negative cooperativity. The advantage in using the index $\kappa$ is that it can be easily measured, yet it ignores all information that can be derived from the shape of $Y(S)$. In particular, this quantifier can be estimated starting from a Klotz plot (see e.g., Fig.~\ref{fig:uno}, panel a)
where the saturation function is shown versus the logarithm of the (free) ligand concentration; in the presence of positive cooperativity this plot yields to a characteristic sigmoidal curve. For the models analyzed here we can estimate $R_{0.9}/R_{0.1}$ directly from the self-consistency equations (\ref{selfC})-(\ref{eq:saturation2}).
Starting from the classical model and, posing $Y=0.9$ and $Y=0.1$ we get, respectively
\begin{eqnarray}
\frac{9}{10} &=& \frac{1}{2} \left \{ 1 + \tanh \left[J \left(2 \times \frac{9}{10} -1\right) + \frac{1}{2} \log(S_{0.9})Ê\right] \right \}, \\
\frac{1}{10} &=& \frac{1}{2} \left \{ 1 + \tanh \left[ J \left(2 \times \frac{1}{10} -1\right) + \frac{1}{2} \log(S_{0.1})Ê\right] \right \},
\end{eqnarray}
and, with some algebra (recalling $2 ~ \textrm{atanh}(x) = \log[(1+x)/(1-x)]$),
%\begin{eqnarray}
%\textrm{atanh} (0.8) &=& Ê0.8 \beta J  + \frac{\beta}{2} \log(S_{0.9}), \\
%\textrm{atanh} (-0.8) &=& Ê-0.8 \beta J  + \frac{\beta}{2} \log(S_{0.1}),
%\end{eqnarray}
\begin{eqnarray}
\nonumber
 \log(S_{0.9}) &=&  2Ê~ \textrm{atanh} \left(\frac{4}{5}\right) -Ê\frac{8}{5}  J  =  \log(9) - \frac{8}{5} J \Rightarrow S_{0.9} = 9 e^{-8J/5}, \\
 \nonumber
\log(S_{0.1}) &=& 2  ~ \textrm{atanh} \left(- \frac{4}{5}\right) + \frac{8}{5} J  = - \log(9) + \frac{8}{5} J \Rightarrow S_{0.1} = \frac{1}{9}
 e^{8J/5},
\end{eqnarray}
that is,
\be
\kappa_{class}= \frac{S_{0.9}}{S_{0.1}} =  81 e^{-16J/5}.
\ee
Of course, when $J=0$ we recover the value $81$, when $J>0$ we get $R_{class} <81$, and when $J<0$ we get $R_{class} >81$.

Repeating analogous calculations for the relativistic model we get
\begin{eqnarray}
\frac{9}{10} &=& \frac{1}{2} \left \{ 1 + \tanh \left[J \frac{2 \times \frac{9}{10} -1}{\sqrt{1 + \left(2 \times \frac{9}{10} -1\right)^2}} + \frac{1}{2} \log(S_{0.9})Ê\right] \right \}, \\
\frac{1}{10} &=& \frac{1}{2} \left \{ 1 + \tanh \left[J \frac{2 \times \frac{1}{10} -1}{\sqrt{1 + \left(2 \times \frac{1}{10} -1\right)^2}} + \frac{1}{2} \log(S_{0.1})Ê\right] \right \},
\end{eqnarray}
and, with some algebra,
\begin{eqnarray}
\nonumber
 \log(S_{0.9}) &=&  2Ê~ \textrm{atanh} \left(\frac{4}{5}\right) -Ê\frac{8}{\sqrt{41}}  J  = \log(9) - \frac{8}{\sqrt{41}} J \Rightarrow S_{0.9} = 9 e^{-8J/\sqrt{41}}, \\
 \nonumber
\log(S_{0.1}) &=& 2  ~ \textrm{atanh} \left(- \frac{4}{5}\right) + \frac{8}{\sqrt{41}} J  = -\log(9) + \frac{8}{\sqrt{41}} J \Rightarrow S_{0.1} = \frac{1}{9}
 e^{8J/\sqrt{41}},
\end{eqnarray}
that is,
\be
\kappa_{rel}= \frac{S_{0.9}}{S_{0.1}} =  81 e^{-16J/\sqrt{41}}.
\ee
Again, one can check that when $J=0$ we recover the value $81$, when $J>0$ we get $\kappa_{rel} <81$, and when $J<0$ we get $\kappa_{rel} >81$.
Also, $\kappa_{rel}/\kappa_{class} = e^{-16J/\sqrt{41} + 16J/5} >1$. This means that, even with this quantifier, when fixing the same coupling constant $J$, the emerging cooperativity is weaker for the relativistic model, as expected.

Next, let us consider the cooperativity quantifier derived from the Scatchard plot.
We recall that this plot is built by showing the behavior of $Y/S$ with respect to $Y$.
In fact, according to the simplest scenario \footnote{This corresponds to the Michaelis-Menten theory and to Clark's theory and it requires a set of simplifying assumptions, among which: the interaction is reversible; all the binding molecules are equivalent and independent; the biological response is proportional to the number of occupied binding sites; the substrate only exists in either a free (i.e., unbound) form or bound to the receptor.}, at equilibrium, one can write
\be \label{eq:equi}
\frac{ S (1-Y)}{Y} = k,
\ee
where $k$ is the proportionality constant between response and occupancy (i.e., it is the ratio between the dissociation and the association constants), and rearranging Eq.~\ref{eq:equi} we have
\be
\frac{Y}{S} = - \frac{Y}{k}  + \frac{1}{k}.
\ee
The previous expression fits the equation of a line for $Y/S$ versus $Y$, whose slope is $-1/k$. The advantages in using the Scatchard plot is that it is a very powerful tool for identifying deviations from the simple model, which, without deviations, is represented by a straight line. In particular, a concave-up curve may indicate the presence of negative cooperativity between binding sites, while a concave-down curve is indicative of positive cooperativity. Also, in the latter case, the maxima occurs at the fractional occupancy $Y^{*}$ which fulfills
\be \label{eq:Scoop}
Y^{*} = \frac{\sigma - 1}{\sigma},
\ee
where $\sigma$ provides another quantifier for cooperativity.
%It should be remarked that the analysis of binding curves by the Scatchard plot is not particularly precise (since the introduction of sophisticated non-linear curve-fitting softwares this method is loosing popularity), yet
%it is a very powerful tool for identifying deviations from the simple model, which without deviations, is represented by a straight line.

%For the models analyzed here we can estimate $n$ directly from the self-consistency equations (\ref{eq:nonrel1})-(\ref{eq:saturation2}).
Starting from the classical model, we can build the function $Y/S$, by first getting $S$ as a function of $Y$ and can be obtained by inverting formula (\ref{selfC}), namely
\be
S(Y) = \exp[ 2 ~ \textrm{atanh}(2Y-1) - 2J(2Y-1)].
\ee
By deriving $Y/S$ with respect to $Y$ we get
\be
\frac{d}{dY} \left( \frac{Y}{S(Y)} \right) = -e^{2J(2Y-1)} [1 - 4J(1-Y)],
\ee
which is identically equal to $-1$ when $J=0$, monotonically decreasing with $Y$ when $J>0$ and monotonically increasing with $Y$ when $J<0$.
The (possible) root therefore provides the extremal point, that is
\be
Y^{*} = \frac{4J-1}{4J},
\ee
and, comparing with Eq.~\ref{eq:Scoop}, we get
\be
\sigma_{class} = 4 J.
\ee

We now repeat analogous calculations for the relativistic model.
First, we get $S$ as a function of $Y$, by inverting formula (\ref{eq:saturation2}), namely
\be
S = \exp \left[ 2 ~ \textrm{atanh}(2Y-1) - 2J \frac{2Y-1}{\sqrt{1+(2Y-1)^2}} \right ].
\ee
By deriving $Y/S$ with respect to $Y$ we get
\be
\frac{d}{dY} \left( \frac{Y}{S} \right) = -e^{\frac{J(2Y-1)}{\sqrt{Y(Y-1) +1/2}} } ~ \left \{ 1 - \frac{4J(1-Y)  }{ [2 - 4Y(1-Y)  ]^{3/2} } \right \},
\ee
which is again identically equal to $-1$ when $J=0$, but it is no longer monotonic when $J \neq 0$ .
More precisely, by studying $d(Y/S)/dY$ we can derive that when $J$ is relatively small, $Y/S$ does not exhibit any extremal points, but there is a flex at intermediate values of $Y$; for intermediate values of $J$ there is a minimum at small values of $Y$ and a maximum at larger values of $Y$; for large values of $J$ there is a maximum.
The extremal points can be found as roots of a $6$-th degree function of $Y$. We can obtain an estimate of the value $Y^{*}$ corresponding to the maximum by recalling $Y \leq 1$ and neglecting high-order terms. In this way we get
\be
Y^{*} \approx \frac{-3 + 2 J^2 - \sqrt{-9 + 26 J^2}}{2 (-9 + J^2)},
\ee
and, comparing with Eq.~\ref{eq:Scoop}, we get
\be
\sigma_{rel} = \frac{2 (-9 + J^2)}{-15 + \sqrt{-9 + 26J^2} }.
\ee

The three plots considered here (i.e., Klotz, Scatchard, and Hill), and the related estimates for the extent of cooperativity, are presented in Fig.~\ref{fig:uno}. In particular, in panel d) we compare the cooperativity quantifiers for several values of $J$: as anticipated, in general, for a given value of $J$, the relativistic model gives rise to a weaker cooperativity.

\begin{figure}
    \includegraphics[width=0.65\textwidth]{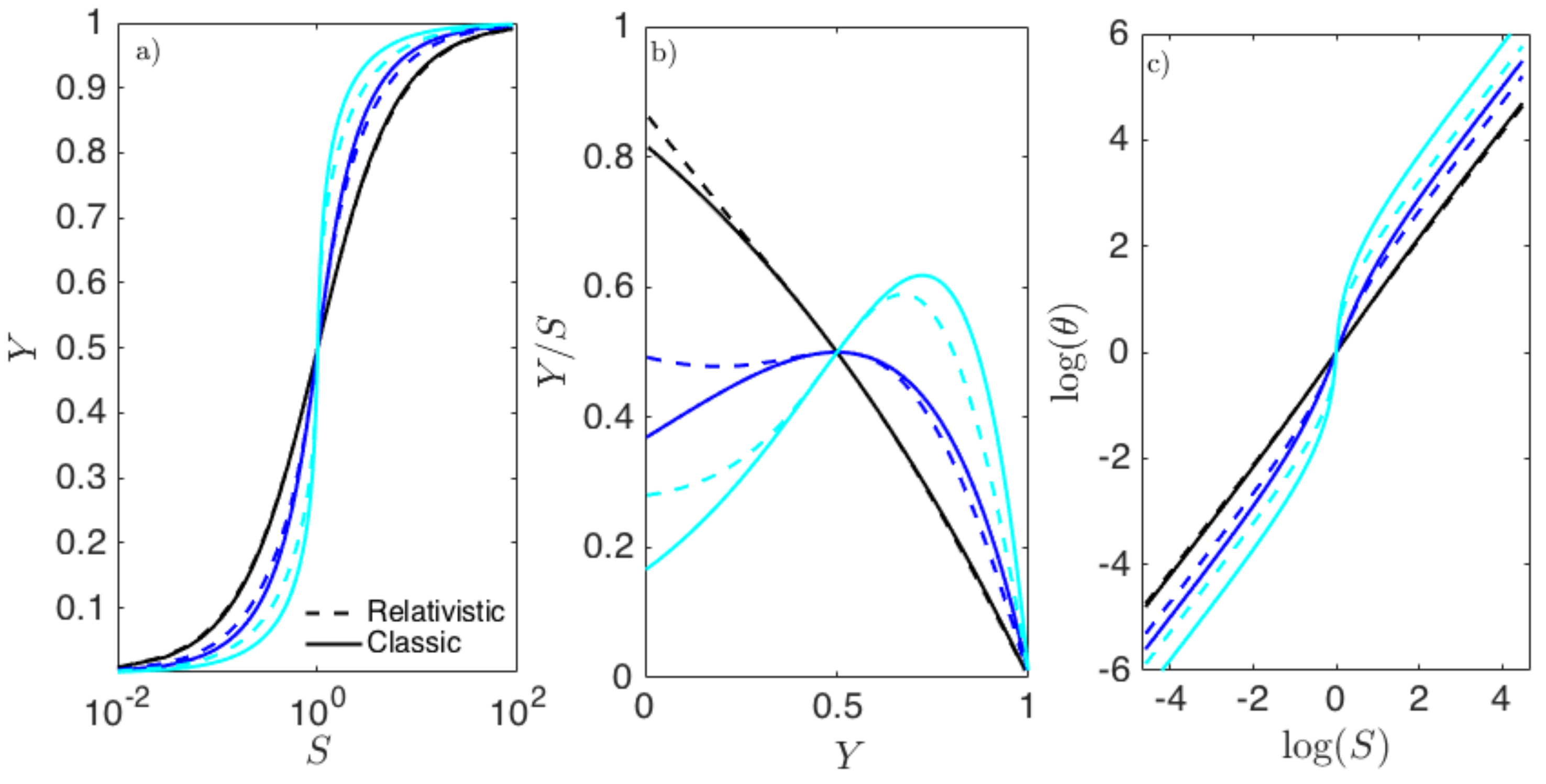}
    \caption{\textbf{The role of the interaction coupling $J$}. We resume the plots presented in panels a)-c) of Fig.~\ref{fig:uno} and we show how they are affected by the interaction coupling $J$. In particular, we compare the outputs for $J=0.1$ (black), $J=0.5$ (blue), and $J=0.9$ (bright blue).  Again, the relativistic model (dashed line) and the classical model (solid line) are compared. Notice that the gap between relativistic and classical model is larger when $J$ is relatively large.}
        \label{fig:due}
\end{figure}

We proceed our analysis by deepening the role of the coupling constant $J$ in the binding curves related to the two models.
In Fig.~\ref{fig:due} we present the Klotz's plot (panel a), the Scatchard plot (panel b), and the Hill plot
(panel c) for the relativistic and the classic models, comparing the outcomes for different values of $J$. As expected, the point corresponding to $S=1$ and $Y=1/2$ is a fixed point in each plot and, in general, the gap between the two models is enhanced when $J$ is larger (i.e., when $J$ is closer to $1$). Also, when $J$ is not too small, the Scatchard plot for the relativistic model displays a flex at small values of $Y$ suggesting that, when the saturation is small, the cooperativity is not truly positive.

In the final part of this section we want to get deeper in the comparison between the classical and the relativistic models.
To this aim, we solved numerically Eq.~\ref{eq:saturation2}, for different values of $S$ and of $J$, getting a set of data $Y(S, J)$. We can think of this set of data as the result of a set of measurements where we collect the saturation value at a given substrate concentration. Now, assuming that in this experiment we have no hints about the underlying cooperative mechanisms, we may apply the formulas for the plain positive cooperativity and infer the value of $J$. More practically, we calculate numerically $Y$ from the relativistic model for different values of $S$ and of the coupling strength, referred to as $J_{rel}$ for clarity. Next, we manipulate the set of data $Y(S, J_{rel})$ by inverting the formula in Eq.~\ref{selfC}: as the value of $S$ is assumed to be known, we can derive the coupling strength, referred to as $J_{class}$, expected within a classical framework. In this way, we can compare the original coupling constant $J_{rel}$ with the inferred one $J_{class}$. We can translate this procedures in formulas as follows:
\begin{eqnarray}\nonumber
J_{class} &=& \frac{\textrm{atanh} (2 Y -1) -\frac{1}{2} \log(S)}{2 Y -1}\\ \nonumber
2 Y -1 &=& \tanh \left[ \frac{J_{rel} (2 Y -1)}{\sqrt{1 + (2 Y -1)^2} }+ \frac{1}{2} \log(S)\right] \\
\Rightarrow J_{class} &=& \frac{J_{rel}}{ \sqrt{1 + (2 Y -1)^2} } \leq J_{rel},
\end{eqnarray}
with equality holding only when $Y = 1/2$.

\begin{figure}
    \includegraphics[width=0.65\textwidth]{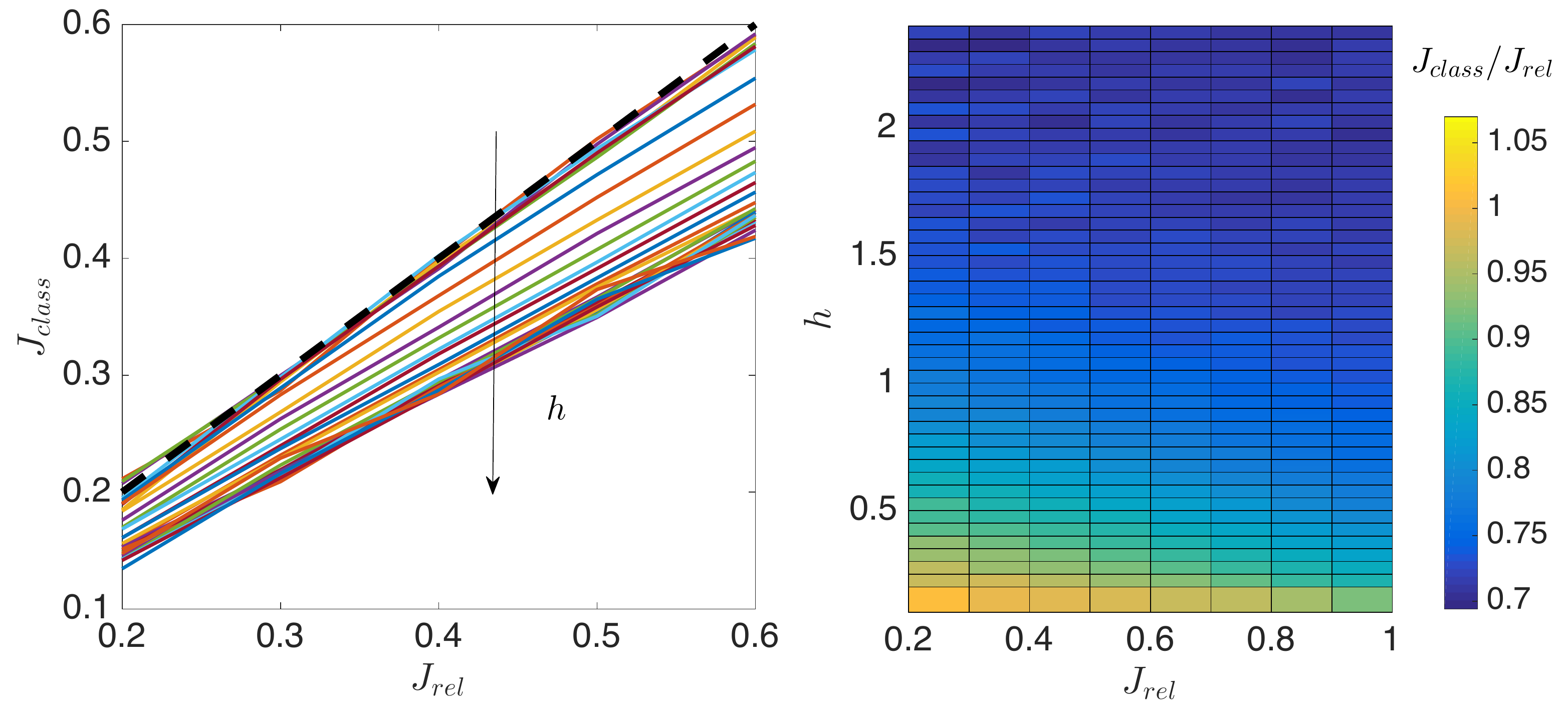}
    \caption{\textbf{Comparison between relativistic and classical model}. We performed numerical experiments where we obtained $Y(S, J_{rel})$ for the relativistic model by solving the self-consistent equation (\ref{eq:saturation2}). From this data we inferred the expected classical coupling $J_{class}$ by inverting the self-consistent equation (\ref{selfC}). We repeated the same operations for several values of $S$ and $J_{rel}.$ In the left panel we show the inferred $J_{class}$ versus the fixed $J_{rel}$: different colors represent different values of $S$ and the identity function is also shown for reference (dashed, black curve). Notice that, in general $J_{class} < J_{rel}$ and the inequality is enhanced as $S$ grows. In the right panel we show a contour plot for the ratio $J_{class}/J_{rel}$ versus $h= \log(S)/2$ and $J_{rel}$. Again, one can notice that, in general, $J_{class}/J_{rel} <1$ and this inequality is enhanced for relatively large values of $S$.}
    \label{fig:tre}
\end{figure}

In Fig.~\ref{fig:tre} (panel a) we plot $J_{class}$ versus $J_{rel}$, for different values of $S$. Notice that the two parameters are related by a linear law, whose slope is smaller than $1$ and decreases with $S$. This confirms that the relativistic model yields to a weak cooperativity. The negative contributions in the relativistic model get more effective when $J_{rel}$ and $S$ are large, as further highlighted in Fig.~\ref{fig:tre} (panel b).

\section{Conclusions}

The rewards in the overall bridge linking Chemical Kinetics and Analytical Mechanics are several, both theoretical and practical, as we briefly comment.
\newline
The former lie in a deeper understanding of the mathematical scaffold for modelling real phenomena: it is far from trivial that the description of chemical/thermodynamical equilibrium is formally the same as the mechanical one. In particular, the self-consistency relation (\ref{eq:self}) that emerges from the thermodynamic principles (in fact, it stems from the requirement of simultaneous entropy maximization and energy minimization) also turns out to be, in the mechanical analogy, a direct consequence of the least action principles  $\delta A(t,x)=0$.
% \begin{equation}
% \label{eq:leastaction}
% \delta A(t,x)=0.
% \end{equation}
This means that the stationary point corresponds to a light perturbation of the evolution of the system in the interval $[0,t]$.
Explicitly, we shift infinitesimally $\braket{M}_{t,x} \rightarrow \braket{M}_{t,x}  + \delta \braket{M}_{t,x} $, then
\begin{equation}
\label{eq:leastaction1}
\begin{split}
0= \delta A(t,x) & = \dfrac{\partial A(t,x)}{\partial \braket{M}_{t,x}} \delta \braket{M}_{t,x}= \\
& =\tanh \left( x-\dfrac{\braket{M}_{t,x} t}{\sqrt{1+\braket{M}_{t,x}^2}} \right) \left( -\dfrac{t \delta \braket{M}_{t,x} }{(1+\braket{M}_{t,x}^2)^{3/2}}\right)+\dfrac{\braket{M}_{t,x} \delta \braket{M}_{t,x} t}{(1+\braket{M}_{t,x}^2)^{3/2}}=0,
 \end{split}
\end{equation}
from which (\ref{eq:self}) is recovered (as usual by setting $t=-J\beta$ and $x=\beta h$), since this holds for all variations $\delta \braket{M}_{t,x} $.
\newline
Even more exciting, still by the theoretical side, is the realization of the {\em complexity} of systems presenting mixed reaction (i.e., where both positive and negative cooperativity are simultaneously at work) and the possible applications in information processing, as we are going to discuss.
\newline
First, let us clarify that in the Literature we speak of {\em complex network} or {\em complex system} with (mainly) two, rather distinct, meanings: in full generality, let us consider  an Hamiltonian as
$$
H(\sigma,J)= \sum_{i<j}J_{ij}\sigma_i\sigma_j
$$
and let us write the two-body coupling matrix as $J_{ij}=A_{ij}W_{ij}$, where $A$ is the adjacency matrix, accounting for the bare topology of the system (its entry $A_{ij}$ is $1$ if there is a link connecting the related nodes $(i,j)$, which are therefore allowed to interact each other, and it is zero otherwise) and $W$ is the weight matrix, accounting for the sign and the magnitude of the links (i.e. the type of interactions among binding sites).
\newline
Dealing with $A$, networks where the topology is very heterogeneous (e.g., the distribution of the number of links stemming from a node has a power-law scaling) are called {\em complex networks}, as it is case for the Barabasi-Albert model \cite{barabasi}.
\newline
Dealing with $W$, networks where the entries of the weight matrix are both positive and negative are termed {\em complex systems}, as the Sherrington-Kirkpatrick model \cite{MPV} for the so called spin-glasses.
\newline
Crucially, spin glasses spontaneously show very general information-processing skills and computational capabilities: for instance Hopfield neural networks \cite{Hopfield} and restricted Boltzmann machines \cite{Hinton}  -key tools in Artificial Intelligence (respectively in {\em neural networks} and {\em machine learning})- are two types of spin-glasses and it is with this last definition of complexity that we now can read the information processing capabilities of the elementary reactions we studied. For a given macromolecule under consideration, we could paste each binding sites on a node and draw the links among nodes that are interacting: if two nodes are correlated (they show positive cooperativity), their relative interaction is positive, while if two nodes are anti-correlated (they show negative cooperativity), their relative interaction is negative.
Dealing with mixed reactions we have to deal with spin glasses and we can thus assess how much information has been processed in a given reaction by evaluating the amount of information processed in its corresponding spin-glass representation, using our bridge. We have already started this investigation in \cite{SciRepBurioni,SciRepKatz,SciRepMoro}.

Finally, from a practical perspective, in the classical limit (i.e., for {\em simple reactions}) we have an explicit expression that directly relates the Hill coefficient $n_H$, which can be measured experimentally, to the interaction coupling $J$ assumed in the model, namely $n_H = 1/(1-J)$. This allows designing specific models and very simple validations (at least at this coarse-grained level) and gives a new computational perspective by which analyze already developed ones (see e.g. \cite{art1,art2,art3,art4,art5}. Then, regarding complex reactions, the puzzling scenario evidenced by Koshland, finally finds out a simple descriptive framework that, crucially, also recovers to the standard cooperative scenario in the proper limit: full coherence among various, apparently antithetic, results is obtained within a unique framework.

\vspace{1cm}
The author declare that there is no conflict of interest regarding the publication of this paper.

\section*{Acknowledgments}
EA and AB are grateful to INdAM-GNFM for partial support via the project AGLIARI2016. AB also acknowledges MIUR via the basal founding for the research (2017-2018) and Salento University for further extra-support.

 %\bibliographystyle{unsrt}
%\bibliography{listb}

\begin{thebibliography}{}

\bibitem{Hinton} D.H. Ackley, G.E. Hinton, T.J. Sejnowski, {\em A learning algorithm for Boltzmann machines}, Cognit. Sci. \textbf{9}.1:147-169, (1985).

\bibitem{Agliari2}  E. Agliari,  et al., {\em Retrieving infinite numbers of patterns in a spin glass model of immune networks},
Europhys. Letts.  \textbf{117}, 28003, (2017).

%\bibitem{Agliari3} E. Agliari,  et al., {\em Immune networks: Multitasking capabilities close to saturation},
%J. Phys. A \textbf{46}, 415003, (2013).

\bibitem{Agliari4} E. Agliari,  et al., {\em Anergy  in self-directed B-cells from a statistical mechanics perspective},
J. Theor. Biol. \textbf{375},  21–31, (2015).

\bibitem{Agliari5} E. Agliari,  et al., {\em A thermodynamical perspective of immune capabilities}, J. Theor. Biol. \textbf{267}, 48, (2011).

\bibitem{SciRepMoro} E. Agliari, et al., {\em Complete integrability of information processing by biochemical ractions},
Sci. Rep. \textbf{6}, 36314, (2016).

\bibitem{SciRepKatz} E. Agliari, et al., {\em Notes on stochastic (bio)-logical gates: computing with allosteric cooperativity},
Sci. Rep. \textbf{5}, 9415, (2015).

%\bibitem{SciRepBisellich} E. Agliari, et al., {\em Cancer driven dynamics of immune cells in a microfluidic enviroment},
%Nature Scientific Reports \textbf{4}, 6639, (2014).

\bibitem{SciRepBurioni} E. Agliari, et al., {\em Collective Behaviours: from biochemical kinetics to electronic circuits},
Sci. Rep. \textbf{3}, 3458,  (2013).

\bibitem{Aldo} E. Agliari, A. Barra, R. Burioni, A. Di Biasio, {\em Mean field cooperativity in chemical kinetics}, Theor. Chem. Acc. \textbf{131}, 1104,  (2012).

\bibitem{Complexity3} P. Andriani, J. Cohen, {\em From exaptation to radical niche construction in biological and technological complex systems}, Complexity \textbf{18}(5),7, (2013).

\bibitem{Angeli} D. Angeli, J.E. Ferrell, E.D. Sontag, {\em Detection of multistability, bifurcations, and hysteresis in a large class of biological
positive-feedback systems}, Proc. Natl. Acad. Sci. USA \textbf{101}(7):1822, (2004).

\bibitem{Tonio1} A. Arsie, P. Lorenzoni, A. Moro, {\em Integrable viscous conservation laws}, Nonlinearity \textbf{28}(6):1859, (2015).

\bibitem{Tonio2} A. Arsie, P. Lorenzoni, A. Moro, {\em On integrable conservation laws}, Proc. R. Soc. A \textbf{471}(2173), 20140124, (2015).

\bibitem{barabasi} A.L. Barabasi, R. Albert, {\em Emergence of scaling in random networks}, Science \textbf{286}(5439), 509, (1999).

\bibitem{Barra1} A. Barra, {\em The Mean Field Ising Model trough Interpolating Techniques}, J. Stat. Phys. \textbf{132}.(5):787, (2008).

\bibitem{Barra-Moro} A. Barra, A. Moro, {\em Exact solution of the van der Waals model in the critical region}, Annals of Physics \textbf{359}:290, (2015).

%\bibitem{Barra3} A. Barra, et al., {\em Notes on P-spin-glass studied via Hamilton-Jacobi and Smooth-Cavity techniques}, J. Math. Phys. \textbf{53}:063304, (2012).

\bibitem{Barra4} A. Barra, A. Di Biasio, F. Guerra, {\em Replica symmetry breaking in mean field spin glasses trough Hamilton-Jacobi technique},
JSTAT P09006, (2010).

\bibitem{Barra5} A. Barra, A. Di Lorenzo, F. Guerra, A. Moro, {\em On quantum and relativistic mechanical analogues in mean field spin models},
Proc. Roy. Soc. A \textbf{470}, 20140589, (2014).


%\bibitem{MicroRNA1} D. Betel, et al., {\em Comprehensive modeling of microRNA targets predicts functional non-conserved and non-canonical sites}, Genome Biol. \textbf{11}.8:R90, (2010).

\bibitem{majani} J.D. Bjorken, S.D. Drell, {\em Relativistic quantum fields}, McGraw-Hill, (1965).

\bibitem{bradshaw} M. Bradshaw, Y. Kubota, T. Meyer, H. Schulman, {\em An ultrasensitive Ca$^{2+}$/calmodulin-dependent protein kinase
  ii-protein phosphatase 1 switch facilitates specificity in postsynaptic calcium signaling},
 Proc. Natl. Acad. Sc. USA \textbf{100}, 10512, (2003).

\bibitem{Crampin} E.J. Crampin, S.Schnell, P.E. Mc Sharry, {\em Mathematical and computational techniques to deduce complex biochemical reaction mechanisms}, Progr. Biophys. $\&$ Molec. Biol. \textbf{86}.1:77, (2004).

\bibitem{Cattoni} D.I. Cattoni, et al., {\em Cooperativity in Binding Processes: New Insights from Phenomenological Modeling}, Plos-One \textbf{10}(12):e0146043, (2015).

\bibitem{chao} L.H. Chao, et al., {\em Intersubunit capture of regulatory segments is a component of
  cooperative camkii activation},  Nature Structural $\&$ Molecular Biology \textbf{17}, 264--272 (2010).

\bibitem{Chen} W.W. Chen, M. Niepel, P.K. Sorger, {\em Classic and contemporary approaches to modeling biochemical reactions}, Genes $\&$ Development \textbf{24}(17):1861, (2010).

\bibitem{Koshland1} R.A. Cook, D.E. Koshland, {\em Positive and Negative Cooperativity in Yeast Glyceraldehyde 3-Phosphate Dehydrogenase}, Biochemistry \textbf{9}(17):3337, (1970).

\bibitem{Koshland2} A. Cornish-Bowden, D.E. Koshland, {\em Diagnostic uses of the Hill (Logit and Nerst) Plots}, J. Mol. Biol. \textbf{95}, 201, (1975).

\bibitem{art5} M. Dougoud, C. Mazza and L. Vinckenbosch, {\em Ultrasensitivity and sharp threshold theorems for multisite systems}, J. Phys. A \textbf{50}(7):075601, (2017).

\bibitem{bookCK} Espenson, J.H. Chemical Kinetics and Reaction Mechanisms, McGraw-Hill (2002).

\bibitem{Mazza} M. Gander, C. Mazza, H. Rummler, {\em Stochastic Gene Expression in Switching Environments}, J. Math, Biol. \textbf{55}:249, (2007).

\bibitem{Barra2} G. Genovese, A. Barra, {\em A mechanical approach to mean field spin models}, J. Math. Phys. \textbf{50}, 053303, (2009).

\bibitem{Giulio} F. Giglio, G. Landolfi, A. Moro, {\em Integrable extended Van der Waals model}, Physica D \textbf{333}, 293, (2016).

\bibitem{Zagrebnov} J.G. Brankov, V.A. Zagrebnov, {\em On the description of the phase transition in the Husimi-Temperley model}, J. Phys. A \textbf{16}.10: 2217, (1983).

\bibitem{Guerra0} F. Guerra, {\em Sum rules for the free energy in the mean field spin glass model}, Fields Inst. Comm. \textbf{30}:161, (2001).

\bibitem{art3} D.G. Hardie, I. P. Salt, S. A. Hawlet, S. P. Davies,  { \em AMP-activated protein kinase: an ultrasensitive system for monitoring cellular energy charge }, J. Chem. Phys. \textbf{338}(3), 717—722, (1999).

\bibitem{storia1} W. Heisenberg, Werner. {\em The revolution in modern science}: Physics and philosophy, (1958).

\bibitem{Hopfield} J.J.  Hopfield, {\em Neural networks and physical systems with emergent collective computational abilities}, Proc. Natl. Acad. Sci. USA  \textbf{79}:2554-2558, (1982).

\bibitem{urka} M. Hucka, et al., {\em The systems biology markup language: a medium for representation and exchange of biochemical network models}, Bioinformatics \textbf{19}.4:524, (2003).

\bibitem{art4} D.C. LaPorte, K.  Walsh, K and D.E. Koshland, {\em The branch point effect. Ultrasensitivity and subsensitivity to metabolic control },  The Journal of biological chemistry, 14068-75, vol. 259, (1984).

\bibitem{Complexity1} C.E. Maldonado, N.A. Gomez-Cruz, {\em Biological hypercomputation: A new research problem in complexity theory}, Complexity \textbf{20}(4):8, (2015).

\bibitem{Katz1} E. Katz, V. Privman, {\em Enzyme-based logic systems for information processing}, Chem. Soc. Rev. \textbf{39}.5:1835, (2010).

\bibitem{art2} D. E. Koshland, G. Némethy, and D. Filmer, {\em  Comparison of Experimental Binding Data and Theoretical Models in Proteins Containing Subunits}, Biochemistry \textbf{5}, 365-385, (1966).

\bibitem{storia2} T.S. Kuhn, {\em The Structure of Scientific Revolutions}, The Physics Teacher \textbf{8}.2:96, (1970).

\bibitem{Science1} H. Kitano,  {\em Systems biology: a brief overview}, Science \textbf{295}.5560:1662, (2002).

\bibitem{MicroRNA2} L. Li, et al., {\em Computational approaches for microRNA studies: a review}, Mammal. Gen. \textbf{21}:1, (2010).

\bibitem{Winfree2} K. Lund, {\em Molecular robots guided by prescriptive landscapes}, Nature \textbf{465}.7295:206, (2010).

\bibitem{mandal} M. Mandal, et al., {\em A glycine-dependent riboswitch that uses cooperative binding to
  control gene expression}, Science \textbf{306}, 275, (2004).

\bibitem{cabibbo} F. Mandl, G. Graham Shaw, {\em Quantum field theory}, John Wiley $\&$ Sons, (2010).

\bibitem{art1} B. Martins, P. Swain, {\em Ultrasensitivity in Phosphorylation-Dephosphorylation Cycles with Little Substrate }, Public Library of Science \textbf{9}, (2013).

\bibitem{bookMazza} C. Mazza and M. Benaim. Stochastic dynamics for systems biology. CRC Press (2014).

\bibitem{gas} J.C. Maxwell, {\em Clerk Maxwell's Kinetic Theory of Gases}, Nature \textbf{8}.(84):122, (1873).

\bibitem{Ricci1} A. Merkoci, et al., {\em Comprehensive analytical chemistry}, Elsevier, (2003).

\bibitem{MPV} M. Mezard, G. Parisi, M.A. Virasoro, {\em Spin glass theory and beyond}, World Scientific (1985).

\bibitem{Michel} A.D. Michel, et al., {\em Direct labelling of the human P2X7 receptor and identification of positive and negative cooperativity of binding}, British Journ. Pharmac. \textbf{151}, 84, (2007).

\bibitem{Neet} K.E. Neet, {\em Cooperativity in Enzyme Function: Equilibrium and Kinetic Aspects}, Methods in Enzyn. \textbf{249}, 519, (1995).

\bibitem{Notides} A. Notides, et al., {\em Positive cooperativity of the estrogen receptor}, Proc. Natl. Acad. Sci. USA \textbf{78}(8):4926, (1981).


\bibitem{Perc1} M. Perc, P. Grigolini, {\em Collective behavior and evolutionary games – An introduction}, Chaos, Sol. $\&$ Fract. \textbf{56}, 1, (2013).

\bibitem{Perc2} M. Perc, M. Marhl, {\em Different types of bursting calcium oscillations in non-excitable cells}, Chaos, Sol. $\&$ Fract. \textbf{18}, 759, (2003).

\bibitem{Complex1} M. Peleg, M.G. Corradini, M.D. Normand, {\em Kinetic models of complex biochemical reactions and biological processes}, Chem. Ing. Techn. \textbf{76}.4:413, (2004).

\bibitem{Winfree1} L. Qian, E. Winfree, {\em Scaling up digital circuit computation with DNA strand displacement cascades}, Science \textbf{332}.6034: 1196, (2011).

\bibitem{Ricci2} F. Ricci, et al., {\em Effect of molecular crowding on the response of an electrochemical DNA sensor}, Langmuir \textbf{23}.12:6827, (2007).

\bibitem{Ricci3} F. Ricci, G. Adornetto, G. Palleschi, {\em A review of experimental aspects of electrochemical immunosensors}, Electrochem. Acta \textbf{84}, 74, (2012).

%\bibitem{Complexity4} P. Schuster, {\em Networks in biology: Handling biological complexity requires novel inputs into network theory}, Complexity \textbf{16}(4), 6, (2011).

\bibitem{Winfree3} G. Seelig, D. Soloveichik, D.Y. Zhang, E. Winfree, {\em Enzyme-free nucleic acid logic circuits}, Science  \textbf{314}(5805):1585, (2006).

\bibitem{Katz2} G. Strack, M. Ornatska, M. Pita, E. Katz, {\em Biocomputing security system: Concatenated enzyme-based logic gates operating as a biomolecular keypad lock}, J.  Amer. Chem. Soc. \textbf{130}(13):4234, (2008).

\bibitem{Thompson} C.J. Thompson, Mathematical Statistical Mechanics, Princeton University Press (1979).

%\bibitem{Complexity2} N.D. Theise, M.C. Kafatos, {\em Complementarity in biological systems: A complexity view}, Complexity \textbf{18}(6):11, (2013).

\bibitem{Valant} C. Valant, P.M. Sexton, A. Christopoulos, {\em Orthosteric-Allosteric bitopic ligands}, Molec. Interv. \textbf{9}(3):125, (2009).

\bibitem{weiss} J.N. Weiss, {\em The Hill equation revised: uses and misuses}, FASEB Journ. \textbf{11}, 835, (1997).

\end{thebibliography}

\end{document}